\documentclass[epj]{svjour}
\usepackage{epsf,epsfig,psfig,float,amssymb,stmaryrd,latexsym}
\usepackage{graphics,psfrag}
\newcommand{\vs}{\vspace{-0.0cm}}
\newcommand{\beq}{\begin{equation}}
\newcommand{\eeq}{\end{equation}}
\newcommand{\beqa}{\begin{eqnarray}}
\newcommand{\eeqa}{\end{eqnarray}}
\newcommand{\nn}{\nonumber \\ }
\newcommand{\bma}{\begin{array}{cc}}
\newcommand{\ema}{\end{array}}

\newcommand{\no}{\nonumber}
\def\3{{\ss}}

\def\vek #1 {\overrightarrow {#1}}
\newcommand{\fet}[1]{\mbox{\boldmath $#1$}}

\begin{document}
\title{Improving the convergence of the chiral expansion for nuclear forces II:  
low phases and the deuteron}

\author{E. Epelbaum \inst{1} \thanks{email: 
                           evgeni.epelbaum@tp2.ruhr-uni-bochum.de}
\and
W. Gl\"ockle \inst{1} \thanks{email:
                           walter.gloeckle@tp2.ruhr-uni-bochum.de}
\and
Ulf-G. Mei{\ss}ner \inst{2} \thanks{email: 
                           meissner@itkp.uni-bonn.de}
}
\institute{ 
  Ruhr-Universit\"at Bochum, Institut f{\"u}r
  Theoretische Physik II, D-44870 Bochum, Germany
\and
Universit\"at Bonn,
Helmholtz--Institut f\"ur Strahlen-- und Kernphysik (Theorie), 
Nu{\ss}allee 14-16,
D-53115 Bonn 
}
\date{Received: date / Revised version: date}
%
\abstract{
Recently we have proposed a new cut--off scheme for  pion
loop integrals in the  two--pion exchange potential. This method allows 
for a consistent implementation of constraints from 
pion--nucleon scattering and has
been successfully applied to peripheral nucleon--nucleon partial waves.
We now consider low partial waves in the 
non--perturbative regime,  where the regularized Lippmann--Schwinger equation 
has to be solved in order to generate the bound and scattering states. 
We observe an  improved description of most of the phase shifts when going
from next-to- to next-to-next-to-leading order in the chiral expansion.
We also  find a good description of the deuteron properties. 
In addition, the new cut--off scheme allows to avoid the presence of unphysical 
deeply bound states. We discuss the cut--off dependence of the four--nucleon
low--energy constants and show that their numerical values can be understood in
terms of resonance saturation. This connects the effective field theory approach 
to boson exchange phenomenology.
}

\authorrunning{E. Epelbaum et al.}
\titlerunning{Improving the convergence of the chiral expansion for nuclear forces II}
\maketitle
\section{Introduction}
\def\theequation{\arabic{section}.\arabic{equation}}
\setcounter{equation}{0}

In Ref.~\cite{EGM03} (which is called I form here on), we have
presented a method to improve the convergence
of the chiral expansion for the nucleon--nucleon (NN) interaction based on spectral function
regularization.  In earlier approaches,  an unphysically 
strong attraction in the isoscalar central part of the  chiral two-pion-exchange (TPE) at 
next-to-next-to-leading (NNLO) order in the chiral expansion was found. 
This is due to the  high--momentum components of the exchanged pions, which appear
when using dimensional  regularization (or equivalent schemes), and which cannot be properly 
treated in the corresponding effective field theory  (EFT).  
Using a cut--off (or spectral function)  regularization 
instead of the dimensional one and taking reasonable values for the momentum 
space cut--off allows to remove spurious short--distance physics associated 
with high--momentum intermediate states 
and to greatly improve the convergence of the chiral expansion. In particular, one
can use without problems the values of the dimension two low-energy constants (LECs)
$c_i$ 
consistent with 
elastic pion--nucleon scattering data.
More precisely, in~I we have
considered the spectral functions obtained from the next-to-leading order (NLO) and 
NNLO TPE contributions and argued that only masses below the chiral symmetry breaking scale 
should be taken into account explicitly in the loop integrals while shorter 
range contributions have to be represented by contact interactions. 
This can be easily implemented by applying a 
cut--off to the spectral functions. We have also proposed 
a simple and convenient way to derive analytic expressions for regularized 
TPE in the momentum space based on the spectral function representation.
  
In~I, we have considered the peripheral partial waves ($l \ge 2$), 
because at NNLO, these are given entirely by one-pion-exchange (OPE) 
and TPE with no free parameters.
We have calculated these phases in Born approximation which should be legitimate
at least for the D-- and higher waves.  The results for the D-- 
and F--waves are still not completely converged at NNLO, but the error of a few~(1)$^\circ$ 
at $E_{\rm lab} = 300$ MeV for the D-- (F--)waves appears reasonable.  
There is no breakdown of the chiral expansion for D--waves beyond $T_{\rm lab} = 50$ MeV and 
for F--waves beyond   $T_{\rm lab} = 150$ MeV as found earlier using dimensional regularization.
In this paper, we apply our method (cut--off regularization, CR, for short) to the low partial 
waves in the  non--perturbative regime,  
where we have to solve the regularized Lippmann--Schwinger equation
to generate the bound and scattering states. As will be demonstrated, there are no deeply bound states 
(for a reasonable range of cut-offs in the spectral function representation of the 
effective potential and regularized Lippmann--Schwinger
equation), and low--energy observables are not affected by CR.

Our manuscript is organized as follows. In section \ref{sec2} we briefly review the 
formalism detailed in~I and discuss the regularization of the Lippmann--Schwinger equation.
In section \ref{sec3} we apply the formalism to {\it np} partial 
waves, with particular emphasis on the low phases (S-- and P--waves). We also discuss the
range parameters in the S--waves and the deuteron properties and compare to the results
obtained in dimensional regularization (DR). The physics behind the values of the 
low--energy constants accompanying the four--nucleon operators is also elucidated.
The summary and conclusions are given in section \ref{sec4}.

\section{Formalism}
\def\theequation{\arabic{section}.\arabic{equation}}
\setcounter{equation}{0}
\label{sec2}

In~I we have calculated the NN potential at NNLO in 
the low--momentum expansion using cut--off regularization in the 
spectral function representation. For completeness, we briefly review
the pertinent formalism here. 
 
The chiral TPE potential up to NNLO can be decomposed into isoscalar 
and isovector central,
spin--spin and tensor components, generically called $W_i (q)$ here.
These functions  $W_i (q)$ can be represented by a continuous superposition
of Yukawa functions (modulo subtractions)
\beq
\label{disp_int}
W_i (q) = \frac{2}{\pi} \int_{2 M_\pi}^\infty \, d \mu \, \mu \frac{\sigma_{i} (\mu)}
{\mu^2 + q^2} \,,
\eeq
where the  $\sigma_i (\mu)$  are the corresponding mass spectra (spectral functions).
Further, $\vec q$ is the momentum transfer in the centre-of-mass system (c.m.s.), 
i.e.~$\vec q = \vec p \, ' - \vec p$, where $\vec p \, '$ and $\vec p$ are final
and initial nucleon momenta, respectively,  and  $q \equiv | \vec q \, |$.
These spectral functions contain
the whole dynamics related to the exchanged two--pion system and can be
obtained from the potential via \cite{Kaiser97}
\beq
\label{spectr:def}
\sigma_i (\mu ) = {\rm Im} \Bigl[ W_i ( 0^+ - i \mu ) \Bigr]\,.
\eeq
The spectral function regularization proposed in~I suppresses the large--$\mu$
contributions to the integrals Eq.~(\ref{disp_int}) 
via a sharp cut--off 
\beq
\sigma_i (\mu ) \to  \theta (\tilde \Lambda - \mu ) \, \sigma_i (\mu )\,,
\eeq
and thus regulates the
short--distance contributions of the TPE in a natural way.

Next, we briefly review the chiral expansion of the NN potential.
The LO potential $V^{(0)}$ is given by 
OPE and two contact interactions, $V^{(0)} = V^{(0)}_{\rm 1 \pi} +
V^{(0)}_{\rm cont}$, with
\beqa
V^{(0)}_{\rm 1 \pi} &=&  -\biggl(\frac{g_A}{2F_\pi}\biggr)^2 \, \fet{\tau}_1 \cdot
\fet{\tau}_2 \, \frac{\vec{\sigma}_1 \cdot\vec{q}\,\vec{\sigma}_2\cdot\vec{q}}
{q^2 + M_\pi^2}\, , \nonumber \\ 
V^{(0)}_{\rm cont} &=& C_S + C_T \, \vec \sigma_1 
\cdot \vec \sigma_2\,,
\eeqa
where $\vec \sigma$ ($\fet \tau $) are nucleon spin (isospin) matrices. 
The NLO corrections are due to two--pion exchange
\beqa
V^{(2)}_{\rm 2 \pi} &=& - \frac{ \fet{\tau}_1 \cdot \fet{\tau}_2 }{384 \pi^2 F_\pi^4}\,
L^{\tilde \Lambda} (q) \, \biggl\{4M_\pi^2 (5g_A^4 - 4g_A^2 -1) \nonumber \\
&& + q^2(23g_A^4 - 10g_A^2 -1)
+ \frac{48 g_A^4 M_\pi^4}{4 M_\pi^2 + q^2} \biggr\}\nn
&& - \frac{3 g_A^4}{64 \pi^2 F_\pi^4} \,L^{\tilde \Lambda} (q)  \, \biggl\{
\vec{\sigma}_1 \cdot\vec{q}\,\vec{\sigma}_2\cdot\vec{q} - q^2 \, 
\vec{\sigma}_1 \cdot\vec{\sigma}_2 \biggr\} \,,\nonumber\\ &&
\eeqa
as well as short--distance contact interactions (local four--fermion terms with two 
derivatives)
\beqa
V^{(2)}_{\rm cont.} &=& C_1 \, q^2 + C_2 \, k^2 +
( C_3 \, q^2 + C_4 \, k^2 ) ( \vec{\sigma}_1 \cdot \vec{\sigma}_2)
\nonumber \\
&+&  iC_5\, \frac{1}{2} \, ( \vec{\sigma}_1 + \vec{\sigma}_2) \cdot ( \vec{q} \times
\vec{k})
+  C_6 \, (\vec{q}\cdot \vec{\sigma}_1 )(\vec{q}\cdot \vec{\sigma}_2 ) \nonumber \\ 
&+& C_7 \, (\vec{k}\cdot \vec{\sigma}_1 )(\vec{k}\cdot \vec{\sigma}_2 )\;.
\eeqa
Here, $\vec k = ( \vec p \, ' + \vec p )/2$, and
the cut--off regularized loop function $L^{\tilde \Lambda} (q)$ reads:
\beqa
\label{def_LA}
L^{\tilde \Lambda} (q) &=& \theta (\tilde \Lambda - 2 M_\pi ) \, \frac{\omega}{2 q} \, 
\ln \frac{\tilde \Lambda^2 \omega^2 + q^2 s^2 + 2 \tilde \Lambda q 
\omega s}{4 M_\pi^2 ( \tilde \Lambda^2 + q^2)}~, \nonumber \\ 
\omega &=& \sqrt{ q^2 + 4 M_\pi^2}~, \nonumber\\
s &=& \sqrt{\tilde \Lambda^2 - 4 M_\pi^2}\,.
\eeqa
The regularized expression for TPE provides an explicit exclusion of the  
short--range components in the spectrum (i.e. those ones with the range 
$r < \tilde \Lambda^{-1}$). 
Furthermore, at NLO one also has a  correction to OPE, It takes the form 
\beq
\label{nlo_ope}
V_{\rm 1 \pi}^{(2)} =  \frac{g_A \, d_{18} \, M_\pi^2}{F_\pi^2} 
\, \fet \tau_1 \cdot \fet \tau_2 \, \frac{(\vec \sigma_1 \cdot \vec q \,) 
(\vec \sigma_2 \cdot \vec q \,)}{q^2 + M_\pi^2}\,,
\eeq 
where the LEC $d_{18}$ is related to the Goldberger--Treiman discrepancy.
Thus, the complete NLO contribution is given by
$V^{(2)} = V_{\rm 1 \pi}^{(2)} + V_{\rm 2 \pi}^{(2)} + V^{(2)}_{\rm cont.}$.
Finally, the cut--off regularized NNLO corrections are represented by the 
subleading TPE potential. It takes the form:
\beqa
\label{ls}
V^{(3)}_{\rm 2 \pi} &=&  -\frac{3g_A^2}{16\pi F_\pi^4}  \biggl\{2M_\pi^2(2c_1 -c_3) -c_3 q^2 \biggr\} 
(2M_\pi^2+q^2) A^{\tilde \Lambda} (q)  \nn
&& {} - \frac{g_A^2}{32\pi F_\pi^4} \,  c_4 (4M_\pi^2 + q^2) A^{\tilde \Lambda}(q)\, 
(\fet{ \tau}_1 \cdot \fet{ \tau}_2 ) \nonumber \\
&&\times \Bigl[ (\vec \sigma_1 \cdot \vec q\,)(\vec \sigma_2 \cdot \vec q\,) 
-q^2 (\vec \sigma_1 \cdot\vec \sigma_2 )\Bigr] \,,
\eeqa
where 
\beq
A^{\tilde \Lambda} (q) = \theta (\tilde \Lambda - 2 M_\pi ) \, \frac{1}{2 q} \, 
\arctan \frac{q ( \tilde \Lambda - 2 M_\pi )}{q^2 + 2 \tilde \Lambda M_\pi}\,.
\eeq

In what follows, we use these values 
for the pion decay constant $F_\pi$,
the pion masses $M_{\pi^\pm}$, $M_{\pi^0}$ and the nucleon mass $m$:
$F_\pi =92.4$ MeV, $M_{\pi^\pm} = 139.570$ MeV,  $M_{\pi^0} = 134.977$ MeV,
$m=938.918$ MeV. In~I we adopted the following values for the nucleon 
axial--vector coupling $g_A$ and the LEC $d_{18}$: $g_A = 1.26$, 
$d_{18}=-0.97$ GeV$^{-2}$.
Alternatively, one can use the larger value $g_A = 1.29$ and completely neglect 
the NLO correction to OPE given in Eq.~(\ref{nlo_ope}) (i.e. set $d_{18} = 0$). 
In this work we will adopt this second possibility. 
Notice, however, that such a replacement is not valid in a general case, 
since the corresponding chiral $g_A$-- and $d_{18}$--vertices 
with three and more pion fields are different.
For the LECs $c_{1,4}$ we adopt the central values  
from the $Q^3$--analysis of the $\pi N$ system  \cite{Paul}:
$c_1=-0.81$ GeV$^{-1}$, $c_4=3.40$ GeV$^{-1}$. For the constant $c_3$ the 
value $c_3=-3.40$ GeV$^{-1}$ is used, which is on the lower side but still
consistent with the results from  
Ref.~\cite{Paul}: $c_3=-4.69 \pm 1.34$ GeV$^{-1}$. This value of the LEC
$c_3$ was found in Ref.~\cite{entem02} to be consistent with empirical 
NN phase shifts as well as the results from dispersion and conventional 
meson theories. Further, the same values of the LECs $c_{1,3,4}$ will 
be used in the upcoming N$^3$LO analysis which will be published separately
\cite{EGMN3LO}. 
Interestingly, similar values for the LEC $c_3$ have been extracted
recently from matching the chiral expansion of the nucleon mass to lattice 
gauge theory results at pion masses between 500 and 800 MeV,
see \cite{Bern03} for more details.
Notice that at NNLO in the chiral expansion of the NN potential
one could, in principle, use the values of the LECs obtained 
in the $Q^2$--analysis of $\pi N$ scattering. This certain freedom in 
choosing the values of $c_{i}$ results in some uncertainty in observables
which might be viewed as an estimation of some higher order effects.  
We would also like to remark that a new {\it np} and {\it pp} partial 
wave analysis of the Nijmegen group \cite{Rent03} leads to 
$c_3 = -4.78 \pm 0.10$ GeV$^{-1}$ and $c_4 = 3.96 \pm 0.22$ GeV$^{-1}$
using $c_1 = -0.76 $ GeV$^{-1}$ as input. These values of 
the LECs $c_{3,4}$ are close to the ones of Ref.~\cite{Paul}.

Using this potential, one can now generate bound and scattering states. For
that, consider the partial--wave projected Lippmann--Schwinger (LS) 
equation for the NN T--matrix:
\beqa
\label{LSeq}
T_{l,\, l'}^{s  j} ( p \, ', \,  p) &=&
V_{l,\, l'}^{s  j} ( p \, ', \,  p) + \sum_{l''} \int \,
\frac{d^3 p''}{(2 \pi)^3} V_{l,\, l''}^{s  j} ( p \, ', \,  p'')
\nonumber \\ && \times
\frac{m}{ p\, ^2 -  (p '')^2 + i \epsilon} 
T_{l'',\, l'}^{s  j} ( p'' , \,  p)\,,
\eeqa
where $V = V^{(0)} +  V^{(2)} +  V^{(3)}$, $m$ is the nucleon mass.
The on--shell S-- and T--matrices are related via
\beq
S_{l, \, l'}^{s  j} (p,  \,  p) = \delta_{l  \, l'} 
- \frac{i}{8 \pi^2}\, p\, m \, T_{l,\, l'}^{s  j} 
( p , \,  p )\,.
\eeq
Contrary to our previous work~I, where we have been interested only in 
peripheral NN scattering and thus calculated the T--matrix perturbatively (i.e. 
keeping only the Born term in Eq.~(\ref{LSeq})), we now have to solve the 
LS equation non--perturbatively. 

Although we have regularized the TPE contributions by cutting off the large--mass 
components in the spectrum (or, equivalently, by explicitly shifting the 
corresponding short--distance components to contact terms), 
the resulting potential still behaves incorrectly
at large $q$ (or equivalently at small $r$).
The effective potential is valid for small values 
of the momentum transfer $q$ and becomes meaningless for $q \gtrsim \Lambda_\chi$.
Moreover, since the potential $V$ grows with increasing momenta $q$, 
the LS equation (\ref{LSeq}) is 
ultraviolet divergent and needs to be regularized.
Following the standard procedure, see e.g.~\cite{EGM2},  
we  introduce an additional cut--off in the LS equation by 
multiplying the potential $V (\vec p, \; \vec p \, ')$ with a regulator function 
$f^\Lambda$,
\beq
\label{pot_reg}
V (\vec p, \; \vec p \, ') \rightarrow f^\Lambda ( p ) \, 
V (\vec p, \; \vec p \, ')\, f^\Lambda (p ' )\,.
\eeq 
In what follows, we use the exponential regulator function 
\beq
f^\Lambda (p ' ) = \exp [- p^6/\Lambda^6 ]~.
\eeq 
Certainly, both cut--offs 
$\tilde \Lambda$ and $\Lambda$ are introduced in order to remove high--momentum
components of the interacting nucleon and pion fields.
The physical meaning and the implementation of the cut--offs
$\tilde \Lambda$ and $\Lambda$ is, however, quite different from each other: 
while the first one removes the short--distance portion of the 
TPE nuclear force, the second one guarantees that the high--momentum 
nucleon states do not contribute to the scattering process. 
One advantage of the method proposed in~I is  that one  can (but does not have to) choose
similar procedures for regulating the spectral functions and the LS equation.
For further discussion of the role and optimal choice of the cut--off  $\Lambda$ in the 
LS equation the reader is referred to Refs.~\cite{Lepage,Geg01}.  

In what follows we will vary  the cut--off  $\Lambda$ in the 
LS equation in the range $450 \ldots 600$ MeV at NLO and $450 \ldots$ 650 MeV 
at NNLO, which is a significantly larger range than in Ref.~\cite{Epe02}. 
The cut--off $\tilde \Lambda$ is varied independently in the 
range $500 \ldots 700$ MeV, which is consistent with the 
variation of $\Lambda$. Notice that in principle, more elegant regularization 
prescriptions, like e.g. lattice regularization, would allow to regularize
pion loop integrals and the Lippmann--Schwinger equation in the same way without
introducing two independent scales $\Lambda$ and $\tilde \Lambda$.

\section{Results}
\def\theequation{\arabic{section}.\arabic{equation}}
\setcounter{equation}{0}
\label{sec3}
For any choice of the cut--offs $\Lambda$ and $\tilde \Lambda$, 
the LECs $C_{S, T}$ and $C_{1 \ldots 7}$ 
are fixed from a fit to the {\it np} S-- and P--waves and the mixing parameter 
$\epsilon_1$ for laboratory energies below 100 MeV. This is the same procedure
as used in Ref.~\cite{EGM2}. In the following,  we 
display and discuss our predictions for the phase shifts at higher energies and 
higher angular momenta as well as for various deuteron properties.  
We also discuss the values of the pertinent low--energy constants (LECs)
and their interpretation in terms of resonance saturation along the lines
of Ref.~\cite{EMGE}.

Before presenting the results we would like to make several comments concerning
the theoretical uncertainty and our way of estimating it. When performing calculations
within chiral EFT up to a certain order $\mathcal{O} (Q^n/\lambda^n)$, where 
$Q \sim M_\pi$ refers
to a generic low--momentum scale (soft scale) and $\lambda$ to the scale at  
which new physics appears (hard scale), the theoretical uncertainty results from 
neglecting the higher order terms and 
is, in general, expected to be of the order 
$\sim (Q/\lambda)^{n+1}$. In particular, one expects for the uncertainty of 
a scattering observable at c.m.s.~momentum $p$ to be of
the order $\sim (\max [ p, \, M_\pi] /\lambda)^{n+1}$.
One should keep in mind that while the hard scale 
$\lambda$ is governed by $\Lambda_\chi \sim M_\rho$ in perturbative calculations
in the $\pi \pi$ and $\pi N$ sectors, where dimensional regularization is 
usually applied and no finite momentum space cut--off is introduced,  
$\lambda \sim \min [ \Lambda, \, \tilde \Lambda , \, M_\rho ] = 450$ MeV should be 
adopted in our case. 

\begin{figure*}[htb]
\vspace{0.5cm}
\centerline{
\psfig{file=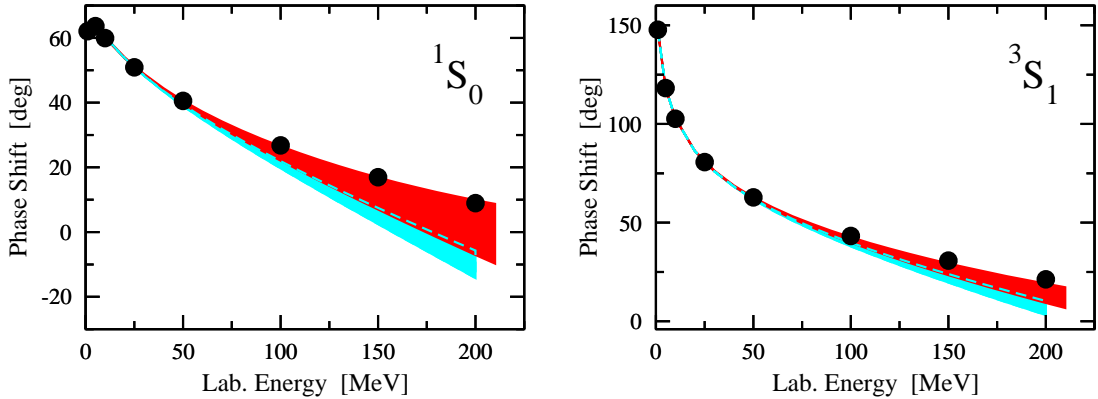,width=15.cm}}
\centerline{\parbox{15.5cm}{
\caption[fig4]{\label{fig1} S--wave NN phase shifts versus
the nucleon laboratory energy. The light (dark) shaded band shows the NLO (NNLO) predictions with 
CR chiral TPE. The cut--off $\Lambda$ in the Lippmann--Schwinger equation is varied in the range 
$\Lambda=450 \ldots 600$ MeV at NLO and $\Lambda=450 \ldots 650$ at NNLO. 
The cut--off $\tilde \Lambda$ in the pion loops is varied independently in the range
$\tilde\Lambda=500 \ldots 700$ MeV at both NLO and NNLO.
The filled circles depict the Nijmegen phase shift analysis (PSA) results \cite{Stoks93}.
}}}
\end{figure*}

Cut--off variation became a common practice to estimate the 
uncertainty and to check consistency of non--perturbative EFT calculations of
few--nucleon systems. Low--energy observables should not depend on the cut--off value 
if all terms
in the EFT expansion are included. In practice, however, calculations are performed at a 
finite order, so that some (small) residual dependence of observables on the cut--off 
remains. One, in general, expects that this cut--off dependence gets weaker when   
higher order terms are included. Thus, at first sight one expects  
narrower bands for scattering observables at NNLO than at NLO for the same 
variation of the cut--off.  This, however, does not hold true for the following reason:
the cut--off dependence at both NLO and NNLO has to be compensated by inclusion of 
the contact interactions (counter terms) of the order $\mathcal{O} (Q^4/\lambda^4)$ 
and higher. The contact interactions appear only at even orders 
$\mathcal{O} (Q^{2l}/\lambda^{2 l})$ 
in the low--momentum expansion while pion exchanges contribute, in general,
at both even and odd orders. Since the same contact terms enter the expression 
for the effective potential at NLO and NNLO, similar cut--off dependence
for observables should be expected  at these orders. Variation of the cut--off
does probably not provide an appropriate estimation of 
theoretical uncertainty at NLO since it does not rely on missing 
$\mathcal{O} (Q^3/\lambda^3)$--terms, but only on 
$\mathcal{O} (Q^4/\lambda^4)$--corrections. 
Notice further that we were only able to vary the cut--off $\Lambda$ in the LS
equation at NLO in the smaller range compared to NNLO, which partially explains  
why the NLO bands in many cases even turn out to be narrower than NNLO ones.
Another reason for that behavior has already been discussed in~I: variation of the spectral 
function cut--off $\tilde \Lambda$ has only small effect at NLO, since the 
leading TPE in most cases provides a very small correction to the LO potential. The corrections
from subleading TPE at NNLO are significantly larger in magnitude (at large $q$),
which leads to larger variation of the potential associated with the spectral  
function regularization, see~I for more details.

\subsection{S--waves}
\def\theequation{\arabic{section}.\arabic{equation}}
\setcounter{equation}{0}

The phase shifts in the $^1S_0$ and $^3S_1$ partial waves are shown in Fig.~\ref{fig1}.
One observes a clear improvement when going from  NLO to NNLO and the description of both 
phases is satisfactory up to the considered energy $E_{\rm lab} = 200$ MeV.
One should keep in mind that this improvement is entirely due to inclusion of the 
subleading TPE potential, since no new contact operators arise at NNLO and thus the 
number of adjustable parameters is the same at NLO and NNLO. 

As expected, the bands at NLO and NNLO are roughly of the same width.
In the $^1S_0$--channel, the band at NNLO is even wider than at NLO, the 
latter, however, does not properly describe the data at larger energies. 
We remind the reader at this point that the bands at NLO underestimate 
the uncertainty of the theory at this order. It is comforting to see that 
the bands at NLO and NNLO overlap and that the Nijmegen values of the S--wave
phase shifts are reproduced at NNLO within the theoretical uncertainty.

\begin{table*}[htb] 
\vspace{1.cm}
\begin{center}
\begin{tabular}{||l||c|c|c||c||}
\hline \hline
{} & {} &  {} & {} & {}\\[-1.5ex]
                & {NLO, CR}                 & {NNLO, CR}                & {NNLO, DR} &{Nijmegen PSA} \\[1ex]
\hline  \hline
{} & {} &  {} & {} & {}\\[-1.5ex]
$a$ [fm]        & $-23.447 \ldots -23.522$  & $-23.497 \ldots -23.689$  & $-$23.936  & $-$23.739  \\[1ex] 
$r$ [fm]        & $2.60 \ldots  2.62$       & $ 2.62  \ldots     2.67$  & 2.73       & 2.68       \\[1ex]
$v_2$ [fm$^3$]  & $-0.46 \ldots  - 0.47$    & $- 0.48 \ldots  - 0.52 $  & $-$0.46    & $-$0.48    \\[1ex]
$v_3$ [fm$^5$]  & $4.3  \ldots   4.4$       & $ 4.0   \ldots     4.2 $  & 3.8        & 4.0        \\[1ex]
$v_4$ [fm$^7$]  & $- 20.7 \ldots  -21.0$    & $-19.9  \ldots   - 20.5$  & $-$19.1    & $-$20.0    \\[1ex]
\hline  \hline
  \end{tabular}
\vspace{0.3cm}
\parbox{13cm}{
\caption{Scattering length and range parameters for the $^1S_0$ partial wave using the
CR NLO and NNLO potential compared to the DR results (with $\Lambda =
1000\,$MeV) and to the Nijmegen phase shift analysis  (PSA). The values 
$v_{2,3,4}$ are based on the {\it np} Nijm II potential and the values of the scattering length 
and the effective range are from Ref.~\cite{Rentm99}.
}\label{tab1}}
\end{center}
\end{table*}

It is also of interest to consider the scattering length and effective range parameters.
The effective range expansion in the S--waves takes the form:
\beq
\label{ere}
p \cot ( \delta ) = - \frac{1}{a} + \frac{1}{2} r \, p^2+ v_2 \, p^4 
 + v_3 \, p^6 + v_4 \, p^8 + \mathcal{O} (p^{10})\,,
\eeq
where $p$ is the nucleon centre-of-mass momentum, $a$ is the scattering length, $r$ is 
the effective range and the $v_i$ are the shape parameters. 
The coefficients in the effective range expansion are governed
by the long--distance physics associated with exchange of pions and thus serve as a good testing 
ground for the convergence of the chiral expansion \cite{Cohen99}. To be specific, let us 
consider the $^1S_0$ partial wave. For each value of the cut--off $\Lambda$ in the Lippmann--Schwinger 
equation one has to determine the values of the LECs $\tilde C_{^1S_0}$,  $C_{^1S_0}$, which 
accompany the (partial--wave projected) contact operators without and with two derivatives, respectively.
These LECs can be fixed, for instance, from the first two coefficients in the effective range expansion,
i.e.~from $a$ and $r$, so that predictions for the $v_i$'s can be made. Alternatively, one can fix them 
from a fit to the phase shift at low energy and then calculate all coefficients $a$, $r$ and $v_i$.
We will adopt this second method in what follows.

\begin{table*}[htb] 
\vspace{1.cm}
\begin{center}
\begin{tabular}{||l||c|c|c||c||}
\hline \hline
{} & {} &  {} & {} & {}\\[-1.5ex]
                & {NLO, CR}                 & {NNLO, CR}                & {NNLO, DR} &{Nijmegen PSA} \\[1ex]
\hline  \hline
{} & {} &  {} & {} & {}\\[-1.5ex]
$a$ [fm]        & $5.429 \ldots 5.433$  & $5.424 \ldots 5.430$    & $5.416$  & $5.420$  \\[1ex] 
$r$ [fm]        & $1.710 \ldots 1.722$  & $1.725   \ldots 1.735$  & $1.756$  & $1.753$  \\[1ex]
$v_2$ [fm$^3$]  & $0.06 \ldots  0.07 $  & $0.04 \ldots  0.05 $    & $0.04$   & $0.04$   \\[1ex]
$v_3$ [fm$^5$]  & $0.77  \ldots 0.81 $  & $ 0.71   \ldots  0.76 $  & $0.67$   & $0.67$   \\[1ex]
$v_4$ [fm$^7$]  & $-4.3 \ldots  -4.4 $  & $-4.1  \ldots   - 4.3$  & $-4.1$   & $-4.0$   \\[1ex]
\hline  \hline
  \end{tabular}
\vspace{0.3cm}
\parbox{13cm}{
\caption{Scattering length and range parameters for the $^3S_1$ partial wave
  using the 
CR NLO and NNLO potential compared to the DR results and to the Nijmegen PSA \cite{Swart95}.
\label{tab2}}
}
\end{center}
\end{table*}

In Table \ref{tab1} we present our results for the effective range coefficients in the $^1S_0$ channel.
Already the NLO results are in a reasonable agreement with the data (as given by the Nijmegen PSA). 
At NNLO we find an improved description for all effective range coefficients. The scattering
length and effective range are still not exactly reproduced at this order if a global fit 
to data is performed. The uncertainty for the scattering length and effective range 
at NNLO resulting from variation of the cut--offs $\Lambda$ and $\tilde \Lambda$  
is of the order of $0.2$ and $0.05$ fm, respectively, which is much larger than the 
deviations from the experimental values ($\sim 0.05$ and $\sim 0.01$ fm). The shape coefficients
are reproduced at NNLO within the theoretical uncertainty. 

We also show the result based on the DR NNLO potential with the cut--off
$\Lambda$ in the Lippmann--Schwinger equation taken as $\Lambda=1000$ MeV. 
Notice that there are two unphysical deeply bound states in each of 
the $^1S_0$ and $^3S_1$--$^3D_1$ channels in that case. 
The results for the effective range coefficients are a bit less precise  
than the ones obtained with CR TPE.

Our results for the effective range parameters in the $^3S_1$ channel are shown in 
Table \ref{tab2}. Similarly to the previously considered case, we observe at NNLO 
an improved description for all coefficients.  
The predicted values of the scattering length and  effective range 
are close to the experimental ones
(within 0.1\% and 1\%, respectively). In the case of the $^3S_1$ channel,
the DR results with $\Lambda=1000$ MeV are slightly more precise than the 
CR ones.

\subsection{P--waves}
\def\theequation{\arabic{section}.\arabic{equation}}
\setcounter{equation}{0}

\begin{figure*}[htb]
\vspace{0.5cm}
\centerline{
\psfig{file=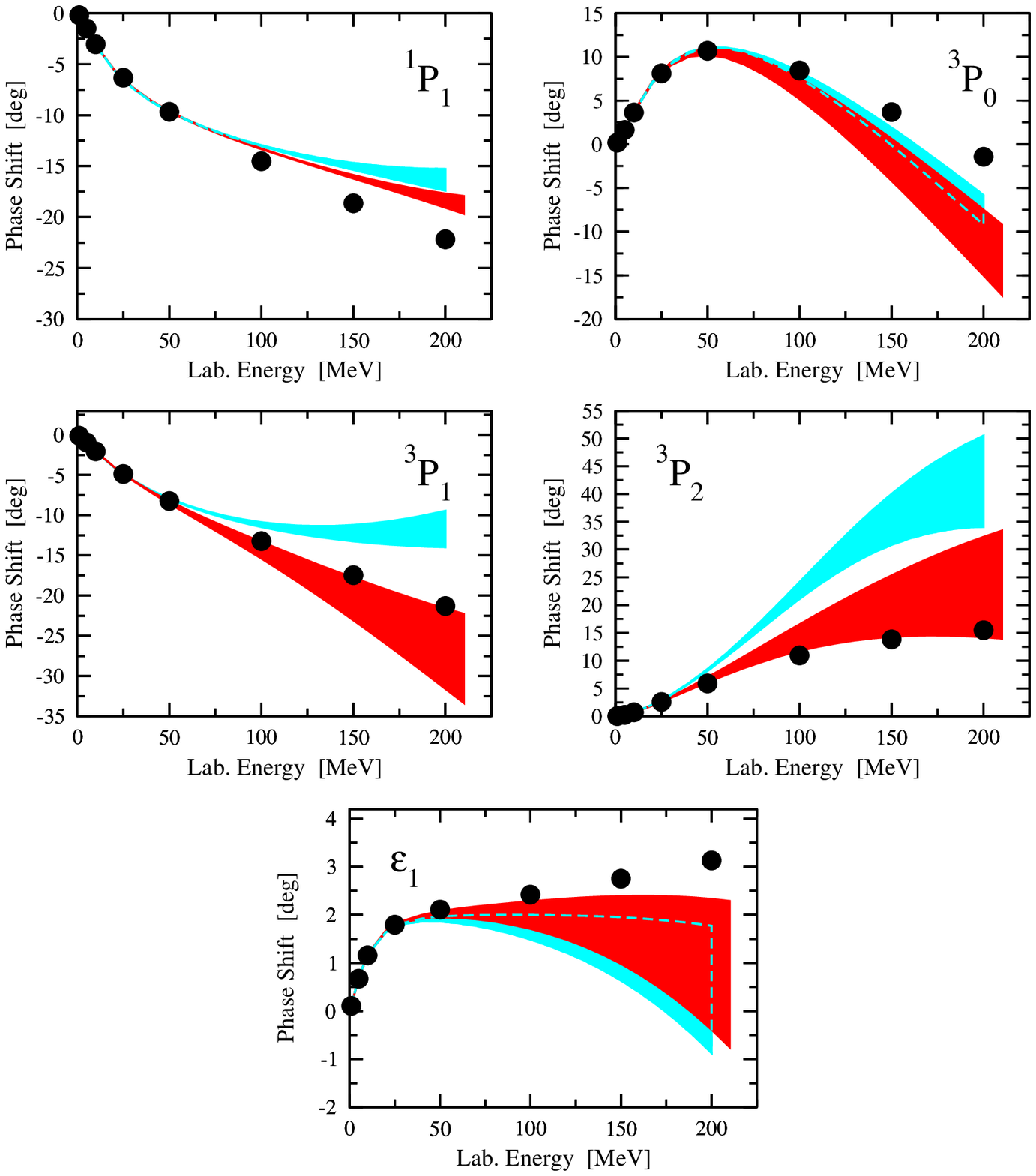,width=15.cm}}
\centerline{\parbox{15.5cm}{
\caption[fig4]{\label{fig2} P--wave NN phase shifts and mixing angle $\epsilon_1$ versus
the nucleon laboratory energy. For notation see Fig.~\ref{fig1}.
}}}
\end{figure*}

Our results for the P--waves and the mixing angle $\epsilon_1$ are shown in Fig.~\ref{fig2}.
While the $^1P_1$, $^3P_1$  and $^3P_2$ phase shifts are visibly 
improved at NNLO compared to the NLO results,
the NNLO results for the  $^3P_0$ phase shift disagree with the data for energies higher than 
$E_{\rm lab} \sim 100$ MeV, where the NLO results are in a better agreement. 
At the moment, we do 
not have an explanation for this disagreement with the data in the  $^3P_0$ channel.
We have checked that this is corrected at N$^3$LO, where a new counter term appears in 
that partial wave. As in the case of S--waves, the bands at NLO and NNLO are 
of a similar width. The theoretical uncertainty in the $^1P_1$ channel is 
probably underestimated by the variation of the cut--offs $\Lambda$, $\tilde \Lambda$. 

In general, our NNLO results 
for the phase shifts based upon CR TPE look similar to the ones of Ref.~\cite{EGM2},
where dimensional regularization has been used to calculate pion loops. 
In the latter case, the TPE potential shows unphysically strong attraction at intermediate 
and short distances, see \cite{Epe02,EGM03} for more details. Although a reasonably good description 
is possible with DR TPE at NNLO, as documented in \cite{EGM2}, unphysical deeply bound states 
arise in D-- and  lower partial waves and one has serious problems with 
the convergence of the chiral expansion. In particular, changing the  value of the cut--off 
in the Lippmann--Schwinger equation clearly leads to a strong variation of the D--wave 
phase shifts, where the potential still turns out to be very strong. The problem with the 
convergence is manifest, since there are no counter terms to compensate this cut--off 
dependence at NNLO.  Using  spectral function regularization in the pion loops as discussed 
in~I we are now able to describe the data equally well as with the DR version and 
in addition:
\begin{itemize}
\item
one can use the same values for the cut--off in the Lippmann--Schwinger equation 
in the LO and NLO 
versions, which are slightly below the mass of the $\rho$--meson,
\item
one does not have spurious deeply bound states,
\item
one has a convergent expansion.
\end{itemize}
Notice that we use here the same values for the LECs $c_{1,4}$ as in Ref.~\cite{EGM2} 
and somewhat smaller in magnitude value for the LEC $c_3$. 
Further, we do not include the (incomplete) set of leading relativistic ($1/m$) corrections 
as done in that paper.

\begin{figure*}[tb]
\centerline{
\psfig{file=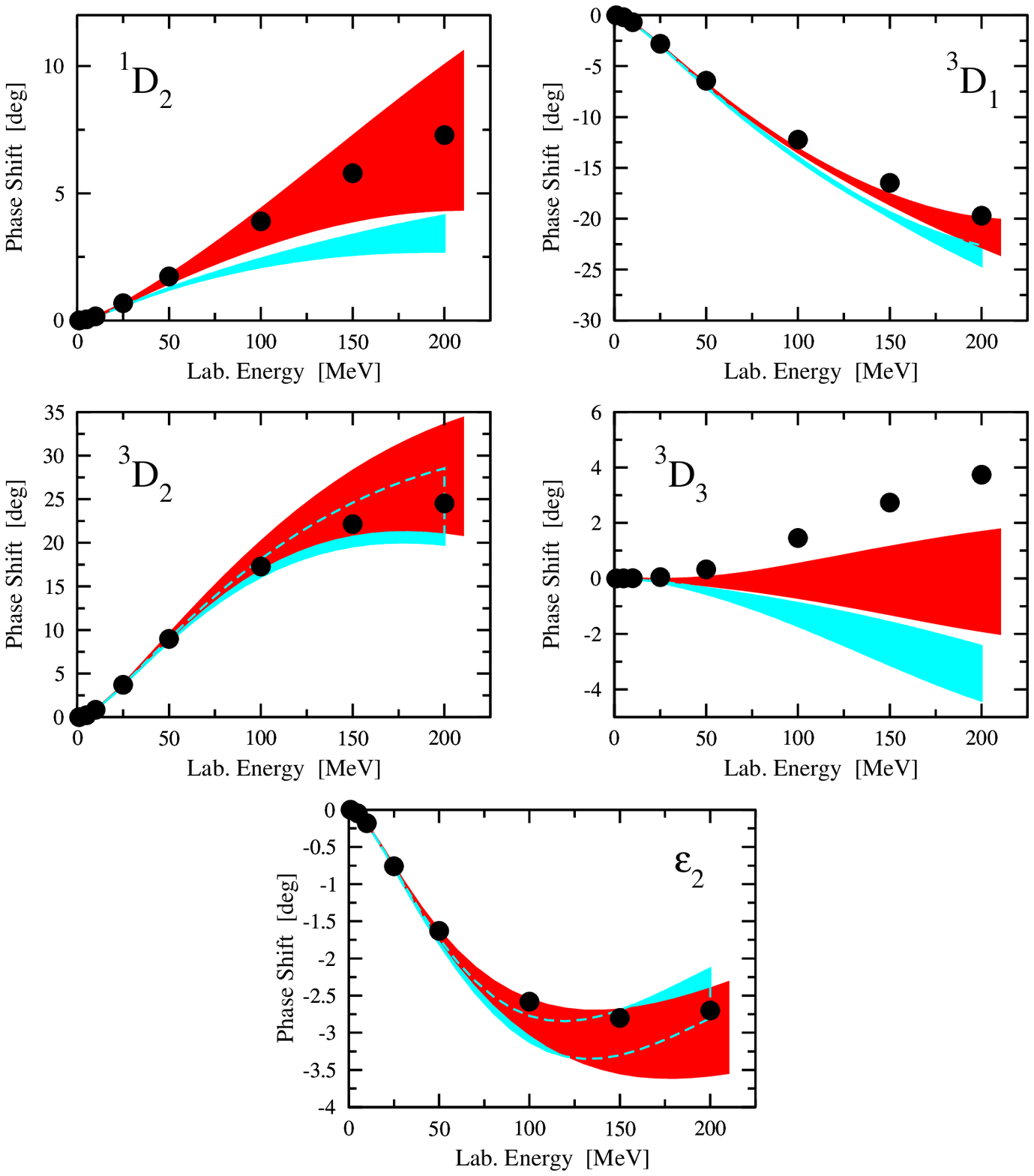,width=15.cm}}
\centerline{\parbox{15.5cm}{
\caption[fig4]{\label{fig3a} D--wave NN phase shifts and mixing angle $\epsilon_2$ versus
the nucleon laboratory energy. For notation see Fig.~\ref{fig1}.
}}}
\end{figure*}

\subsection{D-- and selected higher partial waves}
\def\theequation{\arabic{section}.\arabic{equation}}
\setcounter{equation}{0}

The results for D--waves have already been analyzed in~I
making use of the Born--approximation. In Fig.~\ref{fig3a} we show our 
results for D--waves obtained by solving the LS equation. 
As expected, the results are quite similar to the ones found in~I.
The small differences like e.g. slightly different shape of the 
$^3D_2$ and $\epsilon_2$ phase shifts arise due to introduction of  
the exponential regulator function in Eq.~(\ref{pot_reg}), the 
exact solution of the LS equation as well as due to slightly 
different value of the LEC $c_3$ adopted in the present work. 
Similar results for the phase shifts in the present work and in~I
confirm the high accuracy of the Born approximation in these channels.  

As in the previously considered channels, the bands at NLO and NNLO
are of a comparable width (with exception of the $^1D_2$ partial wave),
and the NNLO results are in a better agreement with the data.  

\begin{figure*}[tb]
\centerline{
\psfig{file=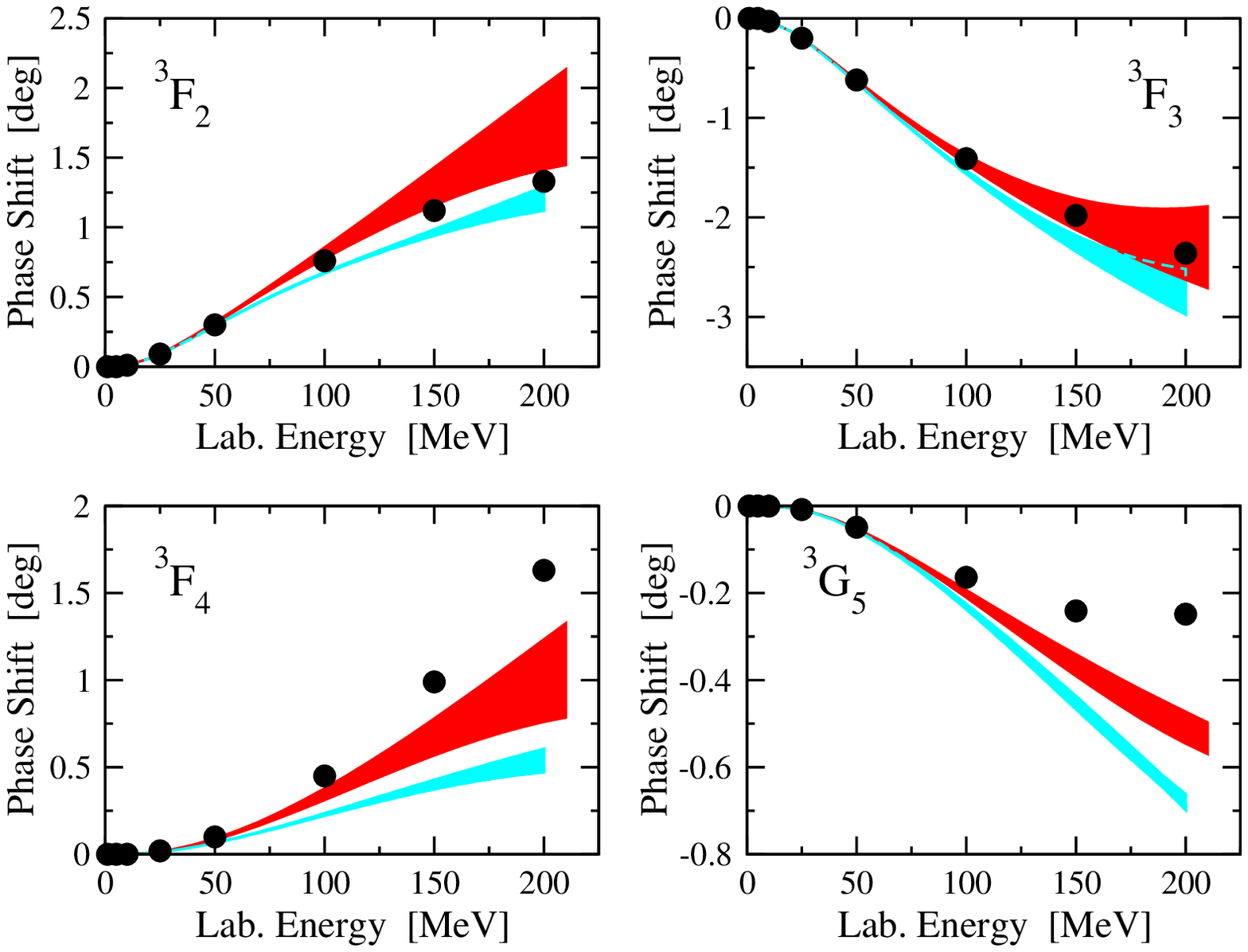,width=15.cm}}
\centerline{\parbox{15.5cm}{
\caption[fig4]{\label{fig3} Selected peripheral NN phase shifts versus
the nucleon laboratory energy. For notation see Fig.~\ref{fig1}.
}}}
\end{figure*}

We  show in Fig.~\ref{fig3} selected higher partial waves,
which also display a very similar behavior to the one observed in~I. 
We remind the reader that a significant disagreement with the data in 
the $^3G_5$ channel at both NLO and NNLO should not be considered 
as a problem because of the exceptionally small value of the phase shift
in this particular channel (more than 10 times smaller in magnitude 
compared to other G--waves).  

\subsection{Deuteron properties}
\def\theequation{\arabic{section}.\arabic{equation}}
\setcounter{equation}{0}

We now turn to the bound state properties. We stress that we do not use the deuteron
binding energy as a fit parameter as it is frequently done but  rather adopt the same
parameters as obtained in the fit to the low phases. In Table~\ref{tab3} we collect the
resulting deuteron properties, in comparison to the DR results with $\Lambda = 1000\,$MeV and
the CD-Bonn potential. First, we note a clear improvement when going 
from NLO to NNLO. In particular
the predicted binding energy deviates by 1\%--1.5\%  at NNLO to be compared with 
$\sim$2\%--2.5\% deviation at NLO. Also visibly improved are the root--mean--square 
matter radius $r_d$ and the asymptotic S--wave normalization strength $A_S$.
For $r_d$, a surprisingly good agreement with the data is observed: the 
NLO (NNLO) prediction deviate from the experimental value by less than 
0.35\% (0.25\%). Deviations from the data for the  
asymptotic S--wave normalization strength $A_S$ of $\sim$2\% ($\sim$1\%) 
at NLO (NNLO) are of the expected size: for a typical deuteron 
observable one expects the uncertainty at NLO (NNLO) to be of the order  
$\sim M_\pi^3 / (450 \mbox{MeV})^3 \sim 3$\% 
($\sim M_\pi^4 / (450 \mbox{MeV})^4 \sim 1$\%). The 
quadrupole moment $Q_d$ is the only deuteron property for which 
our predictions seem to disagree with the observed value by 
somewhat larger amount than expected, namely by $\sim$3.8\%--4.5\% at NLO and 
$\sim$3.8\%--5.2\% at NNLO. It is, however, well known that  $Q_d$
is rather sensitive to short--range physics, see e.g.\cite{Phil01}.  
The deuteron quadrupole moment also turns out to be underpredicted 
by $\sim$4\% in modern potential model calculations. 
Notice further that similar values for $Q_d$ where found in
Ref.~\cite{WM} where DR was used and the
pertinent LECs were fine-tuned to the deuteron binding energy.

\begin{table*}[htb] 
\vspace{1.cm}
\begin{center}
\begin{tabular}{||l||c|c|c||c||c||}
    \hline \hline
{} & {} &  {} & {} & {} & {}\\[-1.5ex]
    & {NLO, CR}  & {NNLO, CR}  & {NNLO, DR} & {CD-Bonn} & {Exp.} \\[1ex]
\hline  \hline
{} & {} &  {} & {} & {} & {}\\[-1.5ex]
$E_d$ [MeV] &   $-2.171 \ldots  -2.186$   & $-2.189 \ldots -2.202$   & $-$2.230  & $-$2.225$^\star$ & $-$2.225 \\[1ex]
$Q_d$ [fm$^2$] &  $0.273 \ldots 0.275$    & $0.271  \ldots  0.275$   & 0.270     & 0.270 & 0.286 \\[1ex]
$\eta$ &     $0.0256 \ldots   0.0257$     & $0.0255 \ldots 0.0256$   & 0.0257    & 0.0255 & 0.0256\\[1ex]
$r_d$ [fm] &   $1.973  \ldots  1.974$     & $1.970  \ldots  1.972$   & 1.970     & 1.966 & 1.967 \\[1ex]
$A_S$ [fm$^{-1/2}$] & $0.868\ldots 0.873$ & $0.874  \ldots  0.879$   & 0.886     & 0.885 & 0.885\\[1ex]
$P_D [\%]$ &     $3.46 \ldots  4.29$      & $3.53   \ldots  4.93 $   & 6.71      & 4.83 &  -- \\[1ex]
    \hline \hline
  \end{tabular}
\vspace{0.3cm}
\parbox{14.8cm}{\caption{Deuteron properties derived from the CR chiral potential
    at NLO and NNLO
    compared to the DR NNLO results of \protect\cite{EGM2},
    one ``realistic'' potential and the data. Here, $E_d$ is the
    binding energy, $Q_d$ the quadrupole moment, $\eta$ the asymptotic
    $D/S$ ratio, $r_d$  the root--mean--square matter radius, $A_S$ the 
    strength of the asymptotic S--wave normalization and $P_D$ the D-state 
    probability. $^\star$ denotes an input quantity. 
\label{tab3}}
}
\end{center}
\end{table*}

The DR NNLO predictions shown in Table \ref{tab3} are in a better (somewhat 
worse) agreement with the data for $A_S$ ($Q_d$) and of the same quality 
for the other observables as the CR NNLO results.

\subsection{Low--energy constants and resonance saturation}
\def\theequation{\arabic{section}.\arabic{equation}}
\setcounter{equation}{0}

We are now in the position to confront the LECs
determined from chiral effective field theory with the highly
successful phenomenological/meson exchange models of the nuclear force
following the lines of Ref.~\cite{EMGE}.
\begin{figure}[tb]
\centerline{\epsfig{file=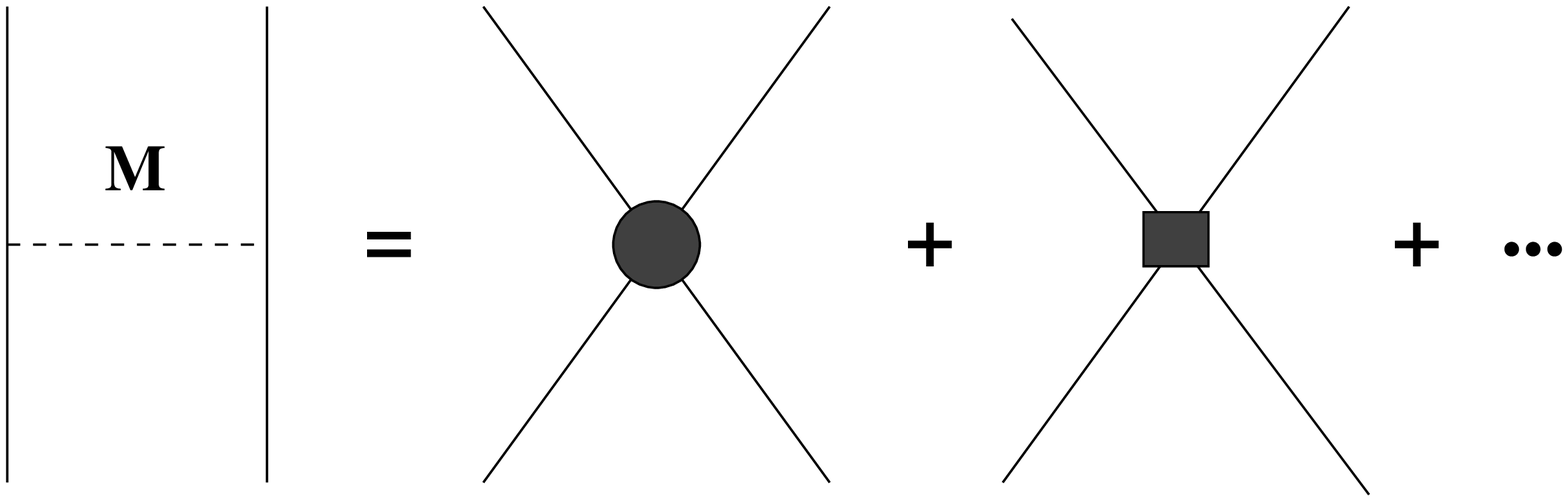,width=2.5in}}

\vspace{0.3cm}
\caption{\protect \small
Expansion of a meson exchange diagram in terms of local four--nucleon
operators. The dashed and solid lines denote the meson $M=\rho, \sigma,
\omega,\ldots$ and the
nucleons, respectively. The blob and the square denote insertions
with zero and two derivatives, in order. The ellipses stands for
operators with more derivatives.}\label{fig:M}
\end{figure}
In this work we will consider the Bonn--B \cite{Machl89} and 
Nijmegen 93 \cite{Sto94} potentials, which are 
genuine one--boson--exchange\\ (OBE) models.
In these models  the long range part of the interaction is given
by OPE (including a pion--nucleon form factor) whereas shorter
distance physics is expressed as a sum over heavier mesons exchange 
contributions:
\beq
V_{\rm NN} = V_\pi + \sum_{M=\sigma, \rho, \ldots} V_M~.
\eeq
Here some mesons can be linked to real resonances (like e.g. the
$\rho$--meson) or are parameterizations of certain physical effects, e.g. the
light scalar--isoscalar $\sigma$--meson is needed to supply the intermediate
range attraction (but it is not a resonance).
The corresponding meson--nucleon vertices are given
in terms of one (or two) coupling constant(s) and corresponding form factor(s),
characterized by some cut--off scale. These form factors are needed to
regularize the potential at small distances (large momenta) but they
should not be given a physical interpretation. 
As depicted in Fig.\ref{fig:M}
for nucleon momentum transfer below the masses of the exchanged mesons,
one can interpret such exchange diagrams as a sum of local operators
with increasing number of derivatives (momentum insertions).
This is explained in detail in Ref.~\cite{EMGE}.
In that work we power expanded the short--range part
of different phenomenological potential models and compared the resulting 
contact operators with the ones in the EFT approach. 
The latter have
to be corrected by adding the corresponding power expanded TPE contributions, 
which are not present in the phenomenological models.   
We have then demonstrated explicitly that the values 
of the LECs $C_i$ determined from various phenomenological OBE 
models are close to the values found in EFT at NLO and NNLO. 
The TPE contributions have been calculated in \cite{EMGE} 
using dimensional regularization. 
In that work we have restricted ourselves to the NNLO version 
with numerically small values of the LECs $c_{3,4}$, 
$c_3=-1.15$, $c_4=1.20$ GeV$^{-1}$, which are not consistent with the $\pi N$ system.
Using the values of these LECs obtained from the $\pi N$ system in 
the TPE potential calculated with dimensional regularization (or equivalent 
schemes) leads, as explained before, to convergence problems due to the strong 
attraction in the central part of the potential. In particular, the 
D--wave phase shifts turn out to be strongly cut--off dependent. 
Further, unphysical  
deeply bound states arise in low partial waves. Obviously, no resonance 
saturation can be established for this NNLO version of the potential, which 
is strongly non--phase--equivalent to the phenomenological OBE models. For example,
the LECs $C_i$ in this NNLO version are typically several times larger in magnitude 
and differ very much from the values at NLO.

\begin{table*}[htb] 
\vspace{1.cm}
\begin{center}
\begin{tabular}{||l||c|c||c|c||c|c||}
    \hline \hline
{} & {} &  {} & {} & {} & {} & {}\\[-1.5ex]
 LEC   & TPE (NLO)  & TPE (NNLO) & $C_i$ (NNLO) & $C_i$ (NNLO) & Bonn B & Nijm-93 \\[1ex]
\hline  \hline
{} & {} &  {} & {} & {} & {} & {}\\[-1.5ex]
$\tilde{C}_{1S0}$  & $-0.004 {{+0.000} \atop {-0.001}}$ & $-0.004 {{+0.000} \atop {-0.001}}$ 
& $-0.117 {{+2.271} \atop {-0.042}}$ & $-0.158 {{+0.178} \atop {-0.004}}$ &  $-0.117$ 
&$-0.061$ \\
{} & {} &  {} & {} & {} & {} & {}\\[-1.ex]
${C}_{1S0}$        & $-0.570 {{+0.036} \atop {-0.022}}$ & $-0.443 {{+0.078} \atop {-0.057}}$ 
& $1.294 {{+2.873} \atop {-0.322}}$  & $1.213 {{+0.408} \atop {-0.084}}$ &  $1.276$
&$1.426$ \\
{} & {} &  {} & {} & {} & {} & {}\\[-1.ex]
$\tilde{C}_{3S1}$  & $0.013 {{+0.001} \atop {-0.000}}$  & $-0.004 {{+0.000} \atop {-0.001}}$ 
& $-0.135 {{+0.025} \atop {-0.021}}$ & $-0.137 {{+0.017} \atop {-0.027}}$ &  $-0.101$
& $-0.014$ \\
{} & {} &  {} & {} & {} & {} & {}\\[-1.ex]
${C}_{3S1}$        & $0.638 {{+0.025} \atop {-0.044}}$  & $-0.443 {{+0.078} \atop {-0.057}}$ 
& $0.231 {{+0.112} \atop {-0.007}}$  & $0.523 {{+0.197} \atop {-0.039}}$  &  $0.660$
& $0.940$ \\
{} & {} &  {} & {} & {} & {} & {}\\[-1.ex]
${C}_{\epsilon 1}$ & $-0.190 {{+0.012} \atop {-0.006}}$ & $0.205 {{+0.024} \atop {-0.035}}$ 
& $-0.325 {{+0.000} \atop {-0.036}}$ & $-0.395 {{+0.007} \atop {-0.072}}$ &  $-0.410$
& $-0.343$ \\
{} & {} &  {} & {} & {} & {} & {}\\[-1.ex]
${C}_{1P1}$        & $-0.067 {{+0.007} \atop {-0.005}}$ & $-0.090 {{+0.013} \atop {-0.009}}$ 
& $0.146 {{+0.005} \atop {-0.010}}$  & $0.126 {{+0.023} \atop {-0.017}}$  &  $0.454$ 
&$0.119$ \\
{} & {} &  {} & {} & {} & {} & {}\\[-1.ex]
${C}_{3P0}$        & $-0.425 {{+0.025} \atop {-0.014}}$ & $0.006 {{+0.003} \atop {-0.003}}$ 
& $0.923 {{+0.142} \atop {-0.103}}$  & $0.920 {{+1.063} \atop {-0.109}}$  &  $0.921$
& $0.802$ \\
{} & {} &  {} & {} & {} & {} & {}\\[-1.ex]
${C}_{3P1}$        & $0.246 {{+0.009} \atop {-0.016}}$  & $0.247 {{+0.032} \atop {-0.044}}$ 
& $-0.260 {{+0.003} \atop {-0.005}}$ & $-0.108 {{+2.364} \atop {-0.176}}$ &  $-0.075$
& $-0.197$ \\
{} & {} &  {} & {} & {} & {} & {}\\[-1.ex]
${C}_{3P2}$        & $-0.022 {{+0.000} \atop {-0.000}}$ & $0.151 {{+0.020} \atop {-0.028}}$ 
& $-0.262 {{+0.032} \atop {-0.073}}$ & $-0.421 {{+0.074} \atop {-0.052}}$ &  $-0.396$
& $-0.467$
  \\[1.5ex]
\hline  \hline
  \end{tabular}
\vspace{0.3cm}
\parbox{13cm}{
\caption{The LECs $C_i$ at NLO and NNLO compared with the results from 
the Bonn B and Nijmegen 93 OBE potential models. Also shown are contributions 
from chiral TPE as explained in text. The $\tilde C_i$ are in 10$^4$ GeV$^{-2}$ 
and the $C_i$ in 10$^4$ GeV$^{-4}$.
}\label{tab:res}}
\end{center}
\end{table*}

\begin{figure}[tb]
\centerline{
\psfig{file=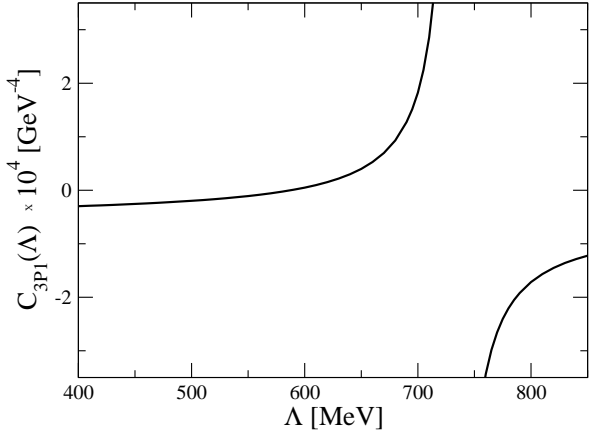,width=8.0cm}}
\caption{\protect \small
``Running'' of the LEC $C_{3P1}$ with the cut--off $\Lambda$ at NNLO.
The cut-off $\tilde \Lambda$ in the spectral function representation
is fixed at our central value $\tilde \Lambda = 600$ MeV.
}\label{fig:3p1}
\end{figure}

We will now demonstrate how resonance saturation works for 
the chiral NN forces at NLO and NNLO introduced above, where 
the new spectral function regularization has been used to derive the TPE 
contributions. Differently to Ref.~\cite{EMGE}, all LECs $c_i$ are now consistent with 
$\pi N$ scattering. In Table \ref{tab:res} we compare the values of the LECs 
$C_i$ at NLO and NNLO with the ones resulting from the OBE models as 
explained before. We remind the reader that the contribution from 
chiral TPE should be accounted for properly in order to allow for a 
meaningful comparison with the OBE models. To achieve that we power 
expand the chiral TPE at NLO and NNLO and identify the corresponding 
contributions to the LECs, which are given analytically in appendix 
\ref{app:TPE}. Notice that differently to \cite{EMGE} these 
contributions depend now on the spectral function cut--off $\tilde \Lambda$.
The second and third columns in Table  \ref{tab:res} show the 
corresponding numerical results at NLO and NNLO for 
the central value $\tilde \Lambda = 600$ MeV, respectively. The indicated
uncertainty refers to the cut--off variation $\tilde \Lambda = 500 \ldots
700$ MeV. The fourth and fifth columns contain the values of the LECs 
$C_i$ at NLO and NNLO, where the just discussed contributions  
from TPE have already been added. 
The numbers are presented for our central 
values of $\Lambda$ and $\tilde \Lambda$, $\Lambda= 550$ MeV and 
$\tilde \Lambda = 600$ MeV and the uncertainties refer to variations 
$\tilde \Lambda = 500 \ldots 700$ MeV and $\Lambda = 450 \ldots 600$
MeV ($\Lambda = 450 \ldots 650$ MeV) at NLO (NNLO). Notice that the 
uncertainties due to the variation of $\Lambda$ and $\tilde \Lambda$ 
are very large in several cases: for $\tilde C_{1S0}$ and $C_{1S0}$
at NLO and for  $C_{3P0}$ and  $C_{3P1}$ at NNLO. Such a strong 
variation in the values of the LECs arises when the cut--off $\Lambda$
in the LS equation becomes too large and one leaves the plateau--region 
for the corresponding LEC $C_i (\Lambda )$. This situation is exemplified
in Fig.\ref{fig:3p1}, where we show the $\Lambda$--dependence of the LEC
$C_{3P1}$ at NNLO for $\tilde \Lambda = 600$ MeV. The   $\Lambda$--dependence
of $C_{3P1}$ is similar to the one of the three--nucleon force
observed in Ref.~\cite{Bed99}.\footnote{The long--range part of the potential 
in that reference behaves like $1/r^2$ at short distances, while the CR chiral 
TPE at NNLO is even more singular and behaves like $1/r^5$, see~I.}
The first branch in Fig.~\ref{fig:3p1} (for $\Lambda \lesssim 730$ MeV) 
corresponds to the case of no deeply bound states. For larger values 
of $\Lambda$ unphysical deeply bound states arise. In that case 
the situation is similar to the DR NNLO analysis of Ref.~\cite{EGM2}.
Clearly, it only makes sense to discuss resonance saturation of the $C_i$'s 
in the plateau--region of the first branch, where they only change modestly
and where the effective potential is at least not strongly non--phase--equivalent 
to the OBE models. Notice that the strong variation of the LECs with the cut--off
did not occur in Ref.~\cite{EMGE} simply because $\Lambda$ was varied in a 
much smaller range and never left the plateau--region.  

The last two columns in Table \ref{tab:res} show the 
LECs as predicted by resonance saturation based upon the Nijmegen 93 
and Bonn B potential models. As in Ref.~\cite{EMGE}, we observe a remarkable
agreement between the LEC values obtained from fit to NN phase shifts 
in the EFT approach and the ones resulting from the OBE models.

\section{Summary and conclusions}
\def\theequation{\arabic{section}.\arabic{equation}}
\setcounter{equation}{0}
\label{sec4}

In this paper we have considered the two-nucleon potential in chiral effective field theory,
making use of the novel method of regularizing the pion loop integrals
introduced in~I. For the low partial waves and the deuteron considered here,
one has to solve the LS equation with the properly regulated potential, cf.
Eq.~(\ref{pot_reg}). One thus has to deal with two different cut--offs, the
first one related to the spectral function regularization (denoted
$\tilde\Lambda$) and the second one related to the regularization of the 
potential in the LS equation  (denoted $\Lambda$). As pointed out, both
of these cut--offs can be chosen in the same range. We obtain the following
results:
\begin{itemize}
\item[1)] We have varied the cut--off  $\Lambda$ in the 
LS equation in the range $450 \ldots 600$ MeV at NLO and $450 \ldots 650$ MeV 
at NNLO, which is a significantly larger range than in Ref.~\cite{Epe02}. 
The cut--off $\tilde \Lambda$ is varied independently in the 
range $500 \ldots 700$ MeV, which is completely consistent with the 
variation of $\Lambda$. 
\item[2)]As shown in Fig.~\ref{fig1} we obtain a modest/satisfactory
 description of the S--waves at NLO/NNLO when fitting the Nijmegen PSA
for energies below 100~MeV. Note that the NNLO result only
differs  from the NLO one by inclusion of subleading two--pion exchange corrections. 
The LECs $c_{1,3,4}$ which enter this subleading TPE contributions
are consistent with the analysis of pion--nucleon scattering in chiral
perturbation theory.
The resulting S--wave scattering lengths, effective ranges and higher
order range parameters come out in good agreement with the ones deduced from
the Nijmegen PSA, cf. Tables~\ref{tab1} and \ref{tab2}. In contrast to earlier
work \cite{EGM2} employing dimensional regularization, 
we have no unphysical deeply bound
states in the S--waves (and any other partial wave).
\item[3)]At NNLO, all P--waves and the mixing parameter $\epsilon_1$ are well
described at NNLO below $E_{\rm lab} \sim 100$ MeV
and improved as compared to their NLO representations
(one parameter per phase at NLO and NNLO), with the exception of the 
$^3P_0$ partial wave. We stress that at NNNLO this deficiency is cured \cite{EGMN3LO}.
\item[4)]The D-- and higher partial  waves come out similar to the results
  obtained in Born approximation in~I. For further discussion, we refer to
  that paper.
\item[5)]The deuteron properties collected in Table~\ref{tab3} are given 
parameter--free. We find a slight improved in the quadrupole moment as
compared to the calculation based on dimensional regularization (as long
as the binding energy is not used as an input parameter).
\item[6)]The theoretical error bars on the various partial waves and the
bound state properties come out consistently with expectations, see in
particular the discussion in the beginning of Section~\ref{sec3}.
\item[7)]We have shown that the numerical values of the four--nucleon LECs
are in good agreement with the ones derived from semi--phenomenological 
boson exchange models once the two--pion--exchange contribution is properly
accounted for, as detailed in the appendix. This strengthens the conclusions
of Ref.~\cite{EMGE} and bridges the gap between the chiral EFT approaches and
more phenomenological models describing the forces between two nucleons.
\end{itemize}

It is now of utmost importance to investigate the next order, that is the NNNLO
corrections, to a achieve a truly accurate description of all important
partial waves. Work along these line is underway \cite{EGMN3LO} (for a first
attempt using dimensional regularization see \cite{EM3}).

\subsection*{Acknowledgments}
\def\theequation{\arabic{section}.\arabic{equation}}

This work is supported in part by the 
Deutsche Forschungsgemeinschaft (E.E.).

\appendix
\section{Reduction of the two--pion exchange contributions}\label{app:TPE}
\def\theequation{\Alph{section}.\arabic{equation}}
\noindent
As stated before, we have to add the contribution of the TPE to the LECs
so as to be able to compare with  boson exchange potentials. 
Expanding the explicit expressions for the renormalized TPE potential given before 
in powers of $\vec{q}$ allows for a
mapping on the spectroscopic LECs (of course, the TPE contains many other
contributions, which are, however, of no relevance for this discussion).
At NLO we get
\beqa
\tilde{C}_{1S0}^{\rm NLO} &=& 
-\frac{1}{3}\tilde{C}_{3S1}^{\rm NLO}
= 18 \, \alpha \, (1 + 4  g_A^2 - 8 g_A^4)  M_\pi^2 \tilde \Lambda^2~, \no\\
C_{1S0}^{\rm NLO} &=&
3  \, \alpha  \, \Big[ (2 + 17 g_A^2 - 88 g_A^4) \tilde \Lambda^2 \no \\
&& \mbox{\hskip 1.9 true cm}{}  - 2 \, 
( 1 + 4  g_A^2 - 8 g_A^4) M_\pi^2 \Big]~, \no\\
C_{3S1}^{\rm NLO} &=&
9  \, \alpha  \, \Big[ -(2 + 17 g_A^2 - 40 g_A^4) \tilde \Lambda^2 \no \\
&& \mbox{\hskip 1.9 true cm}{} + 2 \, 
( 1 + 4  g_A^2 - 8 g_A^4) M_\pi^2 \Big]~, \no\\
{C}_{\epsilon 1}^{\rm NLO} &=&
- 54 \sqrt{2} \, \alpha \, g_A^4 \, \tilde \Lambda^2 ~, \\
C_{1P1}^{\rm NLO} &=&
6  \, \alpha  \, \Big[ (2 + 17 g_A^2 - 16 g_A^4) \tilde \Lambda^2 \no \\
&& \mbox{\hskip 1.9 true cm}{}  - 2 \, 
( 1 + 4  g_A^2 - 8 g_A^4) M_\pi^2 \Big]~, \no\\
C_{3P0}^{\rm NLO} &=&
-2  \, \alpha  \, \Big[ (2 + 17 g_A^2 + 74 g_A^4) \tilde \Lambda^2 \no \\
&& \mbox{\hskip 1.9 true cm}{}  - 2 \, 
( 1 + 4  g_A^2 - 8 g_A^4) M_\pi^2 \Big]~, \no\\
C_{3P1}^{\rm NLO} &=&
2  \, \alpha  \, \Big[ - (2 + 17 g_A^2 - 61 g_A^4) \tilde \Lambda^2 \no \\
&& \mbox{\hskip 1.9 true cm}{}  + 2 \, 
( 1 + 4  g_A^2 - 8 g_A^4) M_\pi^2 \Big]~, \no\\
C_{3P2}^{\rm NLO} &=&
2  \, \alpha  \, \Big[ - (2 + 17 g_A^2 - 7 g_A^4) \tilde \Lambda^2 \no \\
&& \mbox{\hskip 1.9 true cm}{}  + 2 \, 
( 1 + 4  g_A^2 - 8 g_A^4) M_\pi^2 \Big]~, \no
\eeqa
where 
\beq
\alpha = \frac{\sqrt{\tilde \Lambda^2 - 4 M_\pi}}{432 \, F_\pi^4 \, \tilde \Lambda^3 \, \pi}~.
\eeq
The above expressions coincide exactly with the ones given in Ref.~\cite{EMGE} in the 
limit $\tilde \Lambda \to \infty$. Similarly,
we can give the additional TPE NNLO contributions to the various LECs:
\beqa
\tilde{C}_{1S0}^{\rm NNLO} &=&\tilde{C}_{3S1}^{\rm NNLO}
= - 36 \, \beta \, ( 2 c_1 - c_3 )  \, M_\pi^2 \, \tilde \Lambda^2~, \no\\
C_{1S0}^{\rm NNLO} &=&C_{3S1}^{\rm NNLO} =
3 \, \beta \, \Big[ -4 c_4 \tilde \Lambda^2 \no \\
&& \mbox{\hskip 2.35 true cm}{} - 
2 c_1 ( 5 \tilde \Lambda^2 - 2 \tilde \Lambda M_\pi - 4 M_\pi^2)\no \\
&& \mbox{\hskip 2.35 true cm}{}
+ c_3 ( 11 \tilde \Lambda^2 - 2 \tilde \Lambda M_\pi - 4 M_\pi^2 ) \Big]~,\no \\
{C}_{\epsilon 1}^{\rm NNLO} &=&
12 \sqrt{2} \, \beta \, c_4 \, \tilde \Lambda^2 ~,  \\
C_{1P1}^{\rm NNLO} &=&
2 \, \beta \, \Big[ -12 c_4 \tilde \Lambda^2 +  
2 c_1 ( 5 \tilde \Lambda^2 - 2 \tilde \Lambda M_\pi - 4 M_\pi^2)\no \\
&& \mbox{\hskip 2.35 true cm}{} 
- c_3 ( 11 \tilde \Lambda^2 - 2 \tilde \Lambda M_\pi - 4 M_\pi^2 ) \Big]~,\no \\
C_{3P0}^{\rm NNLO} &=&
2 \, \beta \, \Big[ -8 c_4 \tilde \Lambda^2 +  
2 c_1 ( 5 \tilde \Lambda^2 - 2 \tilde \Lambda M_\pi - 4 M_\pi^2)\no \\
&& \mbox{\hskip 2.35 true cm}{} 
- c_3 ( 11 \tilde \Lambda^2 - 2 \tilde \Lambda M_\pi - 4 M_\pi^2 ) \Big]~,\no \\
C_{3P1}^{\rm NNLO} &=&
2 \, \beta \, \Big[ 2 c_4 \tilde \Lambda^2 +  
2 c_1 ( 5 \tilde \Lambda^2 - 2 \tilde \Lambda M_\pi - 4 M_\pi^2)\no \\
&& \mbox{\hskip 2.35 true cm}{} 
- c_3 ( 11 \tilde \Lambda^2 - 2 \tilde \Lambda M_\pi - 4 M_\pi^2 ) \Big]~,\no \\
C_{3P2}^{\rm NNLO} &=&
2 \, \beta \, \Big[ - 2 c_4 \tilde \Lambda^2 +  
2 c_1 ( 5 \tilde \Lambda^2 - 2 \tilde \Lambda M_\pi - 4 M_\pi^2)\no \\
&& \mbox{\hskip 2.35 true cm}{} 
- c_3 ( 11 \tilde \Lambda^2 - 2 \tilde \Lambda M_\pi - 4 M_\pi^2 ) \Big]~,\no 
\eeqa
where 
\beq
\beta = \frac{g_A^2 M_\pi (\tilde \Lambda - 2 M_\pi )}{48 \, F_\pi^4 \, \tilde \Lambda^3}~.
\eeq
These expressions depend on the dimension two LECs $c_{1,3,4}$ as discussed
before and coincide with the ones given in Ref.~\cite{EMGE} in the limit 
$\tilde \Lambda \to \infty$ modulo $1/m$--corrections, which are not considered in 
the present work. 

Finally, it should be kept in mind that we use here the same notation as in our previous 
work \cite{EGM2}, according to which the contact terms that are nonanalytic in the pion mass and 
result from TPE, see e.g.~I, are not shown explicitly. In practical applications 
to the NN system 
at fixed value of $M_\pi$ such terms cannot be disentangled from the zero--range counter terms
(contact interactions from the Lagrangian) and therefore do not need to be treated separately. 
A more precise treatment is required if the pion mass dependence of the nucleon force is 
studied \cite{EMG02,BScl1,BScl2}.

\end{document}